\documentclass[fleqn]{2017SCGE}
\newcommand{\mnras}{Mon. Not. R. Astron. Soc.}
\newcommand{\apj}{Astrophys. J.}
\newcommand{\ana}{Astron. Astropys.}
\newcommand{\revmp}{Rev. Mod. Phys.}
\newcommand{\pasa}{Publ. Astron. Soc. Aust.}
\newcommand{\astrp}{Astron. Rep+.}
\newcommand{\annms}{Ann. Math. Stat.}
\newcommand{\scich}{Sci. China-Phys. Mech. Astron.}

\begin{document}

\ensubject{subject}

\ArticleType{Article}
\SpecialTopic{SPECIAL TOPIC: }%
\Year{2019}
\Month{...}
\Vol{.}
\No{1}
\DOI{...}
\ArtNo{000000}
\ReceiveDate{...}
\AcceptDate{...}

\title{Measuring clock jumps using pulsar timing}

\author[1,3,4]{ZhiXuan Li}{}
\author[2]{KeJia Lee}{{kjlee@pku.edu.cn}}
\author[2]{Ricardo Nicolaos Caballero}{}
\author[1,3,4]{YongHua Xu}{}
\author[1,4]{\\LongFei Hao}{}
\author[1,4]{Min Wang}{}
\author[1,4]{JianCheng Wang}{}

\AuthorMark{Li Z. -X, Lee K. -J, et al.}

\AuthorCitation{Li Z. -X, Lee K. -J, et al}

\address[1]{Yunnan Observatories, Chinese Academy of Sciences, Kunming 650216, China}
\address[2]{Kavli Institute for Astronomy and Astrophysics, Peking
University, Beijing 100871, China}
\address[3]{University of Chinese Academy of Sciences, 19A Yuquan Road, Shijingshan District, Beijing 100049, China}
\address[4]{Key Laboratory for the Structure and Evolution of Celestial Objects, Chinese Academy of Sciences, Kunming 650216, China}



\abstract{In this paper, we investigate the statistical signal-processing 
algorithm to measure the instant local clock jump from the timing data of 
multiple pulsars. Our algorithm is based on the framework of Bayesian 
statistics. In order to make the Bayesian algorithm applicable with limited 
computational resources, we dedicated our efforts to the analytic marginalization 
of irrelevant parameters. We found that the widely used parameter for 
pulsar timing systematics, the `Efac' parameter, can be analytically 
marginalized. This reduces the Gaussian likelihood to a function very similar 
to the Student's $t$-distribution. Our iterative method to solve the maximum 
likelihood estimator is also explained in the paper. Using pulsar timing data 
from the Yunnan Kunming 40m radio telescope, we demonstrate the application of the 
method, where 80-ns level precision for the clock jump can be achieved. Such a 
precision is comparable to that of current commercial time transferring service 
using satellites. We expect that the current method could help developing the 
autonomous pulsar time scale.}



\maketitle


\begin{multicols}{2}
\section{Introduction}\label{sec:1}

Pulsars are stable celestial clocks, particularly, the long-term
stability of millisecond pulsars (MSPs) is comparable to that of the 
current
atomic time-scales \cite{MT97, HL11}.
On a short time-scale, the noise in 
pulsar timing signals
becomes signal-to-noise ratio limited, and it is widely accepted that the atomic 
clock is more stable \cite{HL11}.  At first glance, 
pulsar timing would not help much, if one focuses on
the short term timing stability. Indeed, most of the previous studies
aimed at developing a pulsar time-scale with long-term stability \cite{R08, GP11, 
HCM12, CLL16, RF18}, 
while the short-term stability is not extensively investigated.

The traditional local time-frequency sources based on the 
\Authorfootnote atomic clock to generate the frequency standard, 
GPS technology have two major components: (i) a local 
and (ii) a GPS receiver as a servo to regularize the long-term timing stability 
\cite{kaplan05}. The atomic frequency standard easily reaches a daily Allan 
variance of $10^{-13}$, while the GPS system performs rather well in time 
broadcasting, and the precision of 10 ns or better can be achieved using the 
commercial systems \cite{mic2015}. Clearly, it is unreasonable to expect that 
pulsar timing will surpass such high stability on short-timescales. 

However, the above mentioned stability of the atomic frequency standard is based on 
the assumption that the atomic frequency standard is well behaved. In practice, 
due to environmental variations, frequency standard tuning, or other factors (e.g. 
accidentally changing clock configuration, replacing cables, or interference in 
the GPS receiver), the local time standard sometimes shows unexpected jitter.  
In most of the cases, the time signal jumps for a few micro- to milli-seconds. This 
is a rather common phenomenon, and appears nearly at all astronomical
observatories (for an example, see \cite{LKG16}). Clearly, the very first step to construct a standalone pulsar 
timescale aiming at real scenario applications would be the identification of 
those signal jitters and related corrections.

In this paper, we propose to use the timing data of multiple pulsars to diagnose 
and correct such local time jitter. This is very similar to a pulsar timing 
array (PTA) experiment, which primarily aims at detecting gravitational waves at the nano-Hz band 
\cite{FB90}.  We introduce the algorithm to compute the local clock jumps from 
pulsar timing data in Section 2. In Section 3, we demonstrate the algorithm 
using pulsar timing data from the Kunming 40m (KM40m) radio telescope. 
The conclusions and discussions are made in Section 4.

\section{Waveform estimation for clock jumps}\label{sec:2}

In the common pulsar timing practice, one measures the TOA at the telescope 
site. The TOA is then converted to the TOA as seen by an observer at the 
pulsar's co-rotating reference frame. During the conversion, one accounts for 
effects such as Earth motion in the Solar system, electromagnetic wave 
dispersion, pulsar motion with respect to the Solar-system barycenter. The 
orbital dynamics of the pulsar also need to be corrected, if it is in a binary 
system. The next step is to compute the pulsar timing residuals. Here, an 
integer phase is subtracted from the pulsar-frame TOA, and the difference 
between the TOA and the modeled integer phase is the timing residual.  
All the unmodeled effects, such as clock jitter, reside in these timing 
residuals. We refer the interested readers \cite{LK12} and references therein for 
an extensive description of pulsar timing techniques.

The clock jitter or jump introduces a \emph{common mode} in all pulsars’ timing 
data, i.e. all the TOAs will be late for a given value at the same epoch, if the 
clock ticking leaps forward. Such a common mode is very different from other 
possible noise signals, e.g. pulsar intrinsic noise will not be correlated among 
pulsars. The clock-induced common mode is identical for all pulsars. We need to 
identify this identical signal and estimate the waveform in order to compute 
the clock jump. 

Our algorithm to extract the clock jump is based on the maximum likelihood 
estimator (MLE). For the waveform estimation purpose, the MLE was proven to be 
asymptotically optimal \cite{wilks1938large}. Such a property guarantees that the 
error of waveform estimation approaches the best possible values when the 
signal-to-noise ratio becomes large.

Clock jitter happens on timescales usually less than a few minutes. We can use
pulsar timing data around the jitter epoch to determine the clock jitter 
amplitude. Neglecting the red noise contribution for short timescales\footnote{We 
can neglect the red noise, if its power is much lower than measurement error.  
For data span of less than one year, we can neglect red noise of most PTA MSPs 
such as that measured by \cite{CLL16} for long-term datasets.}, 
the timing residuals contain mainly two parts, the clock jitter and 
measurement error, i.e.  \begin{equation}
	R_{i,j}=s_{j}+n_{i,j}\,,	\label{eq:sig}
\end{equation}
where we denote the timing residual of the $i$-th pulsar at the $j$-th epoch as 
$R_{i,j}$, $s_{i, j}$ is the waveform of the clock jitter and $n_{i,j}$ is the 
measurement error, following the same convention for the indices as for the timing residual ($R$). 
The noise ($n_{i,j}$) is a zero mean Gaussian random variable. In this paper, we assume 
that there are extra systematics in determining the errorbar of each TOA. We 
summarize the systematics using the `Efac' parameters similar to previous studies 
\cite{HEM06b}. The likelihood for the timing residual thus becomes
\begin{equation}
	f(R_{i,j})\propto  \prod_{i=1}^{N}\prod_{j=1}^{M_i} \frac{1} {\eta_{i} 
	\sigma_{i,j}} e^{-\frac{1}{2} \left(\frac{R_{i,j}-s_{i, j}-\mu_{i} }{\eta _{i} 
	\sigma_{i,j} } \right)^2}\,.
	\label{eq:lik}
\end{equation}
Here, the function $f$ is the likelihood function, i.e. the probability density for 
the given data ($R_{i,j}$). $N$ and $M_i$ are the total number of pulsars and 
epochs for the $i$-th pulsar. $\mu_{i}$ is the local mean of the residual, 
$\sigma_{i,j}$ is the errorbar of each data point.  The `Efac' parameter 
$\eta_i$ is defined for each pulsar to include the minimum practical modeling 
for the measurement of systematics.

Focusing on the clock jitter, the signal will be identical for all pulsars. We 
model the clock jitter using a step function of the form
\begin{equation}
	s_{i, j}= \left\{\begin{array}{l} s_0, \textrm{for } T_{i,j}>t_0\,, \\
		0, \textrm{ for alternatives}\,,
	\end{array}\right.
	\label{eq:jmpwave}
\end{equation}
where $s_0$ is the amplitude of the clock jitter, which allows for both positive 
and negative values corresponding to the two possible ways of clock jitter, i.e. delay or advance.  
The time epoch of the data points is described by $T_{i,j}$. 

We can now split the likelihood into a multiplication of two independence parts, 
the one before the clock jitter and the one after it, i.e.
\begin{equation}
	f(R_{i,j})\propto  \prod_{i=1}^{N}\prod_{j=1}^{\rm before } \frac{1} {\eta_{i} 
	\sigma_{i,j}} e^{-\frac{1}{2} \left(\frac{R_{i,j}-\mu_{i} }{\eta _{i} 
	\sigma_{i,j} } \right)^2}  \prod_{i=1}^{N}\prod_{j={\rm after}}^{M_i } 
	\frac{1} {\eta_{i} \sigma_{i,j}} e^{-\frac{1}{2} 
	\left(\frac{R_{i,j}-s_0-\mu_{i} }{\eta _{i} \sigma_{i,j} } \right)^2}\,.
	\label{eq:lik2}
\end{equation}

By inserting Equation~\ref{eq:jmpwave} into Equation~\ref{eq:lik2}, one can see that the 
likelihood is controlled by $2+2N$ parameters. There are two parameters for the 
clock jitter, the event's amplitude and epoch ($s_0$ and $t_0$), $N$ pulsar 
`Efac' parameters ($\eta_i$) and finally $N$ local mean values ($\mu_i$). 
The MLEs for all the parameters 
are derived according to
\begin{equation}
	{s_0, t_0, \mu_i, \eta_i}={\rm argmax}[f(R_{i,j}) ]\,.
	\label{eq:maxlik}
\end{equation}
The statistical inference seems to be straightforward, e.g. one can perform
Monte-Carlo Bayesian inference with the likelihood function 
(Equation~\ref{eq:lik2}) to measure all $2+2N$
parameters and related errors. 

However, for the current problem, such a full Monte-Carlo Bayesian method is not 
ideal. It has a very high computational cost due to the potentially large 
number of parameters. On the other hand, the Bayesian method would perform the inference 
on all parameters, for some of which we are not interested in their values, e.g.  $\eta_i$ and $\mu_i$ 
are parameters for individual pulsars. Furthermore, we only need 
to measure the epoch of jitter to the precision of observation cadence due to 
the sparseness of pulsar timing observations. The requirement for jitter-epoch 
estimation is rather rough. As we will show in the following paragraphs, it is 
possible to compute the MLE in a very computationally efficient way, if we perform 
a one-dimensional search for the jitter epoch ($t_0$). That is, we can compute 
the MLE and error for the clock jump amplitude using a direct method, at each 
trial jitter epoch.  

Since `Efac' parameters are not required, we can marginalize the them by 
integrating the likelihood function ($f$) under \emph{Bayesian prior}, and the 
reduced likelihood ($f'$) is then defined as
\begin{equation}
	f'(R_{i,j})\equiv\int f(R_{i,j}) p(\eta)\,{\rm d} \eta\,,\label{eq:likred}
\end{equation}
where function $p(\eta)$ is the \emph{prior} of $\eta$. The least informative 
prior will be $p(\eta)\propto1/\eta$, i.e. when the $\eta$ probability distributes uniformly in 
logarithmic space \cite{gregory2005bayesian}. Inserting Equation~\ref{eq:lik2} 
into Equation~\ref{eq:likred}, one can show that \footnote{We have used integral 
of
$\int_0^{\infty} \eta^{-n-1} \exp\left[-a/(2\eta^2)\right]\,d\eta =2^{n/2-1} 
a^{-n/2} \Gamma(n/2)$ \cite{gr14}, where $\Gamma$ is the Gamma-fucntion.}
\begin{equation}
	f'(R_{i,j})\propto  \prod_{i=1}^{N}\left[\sum_{j=1}^{\rm before } 
	\left(\frac{R_{i,j}-\mu_{i} }{\sigma_{i,j} } \right)^2
	+\sum_{j={\rm after}}^{M_i}  \left(\frac{R_{i,j}-\mu_{i} -s_0}{ \sigma_{i,j} } 
	\right)^2\right]^{-M_i/2}\,.
	\label{eq:likred2}
\end{equation}
The above result is very similar to the Student's $t$-distribution with $M-1$ degrees
of freedom, which describes the mean of a normally distributed population with 
finite sample size and unknown standard deviation. 
Note that for any time epoch between two available data points, 
the data defining `before' and `after' are the same, and as such
for all time epochs between these points the likelihood value will be the same.
In this way, the best measurement precision for the epoch of clock jump is the 
length between the two nearest data points containing the jump. This becomes
more apparent in case where the jump is associated with data gaps longer than the average data cadence,
as we will also see in Section 3.

The MLE for $\mu_i$ can be readily computed from $\partial f'/\partial \mu_i=0$, 
which leads to
\begin{equation}
	\hat{\mu_i}=\frac{\sum_{j=1}^{\rm before} R_{i,j} \sigma_{i,j}^{-2}} { 
	\sum_{j=1}^{\rm before}\sigma_{i,j}^{-2}}\,.
	\label{eq:estimu}
\end{equation}
Here the hat symbol $\hat{X}$ stands for the estimator for any given parameter, $X$.
The condition for the MLE for $s_0$ is $\partial f'/\partial s_0=0$, which gives 
an implicit expression
\begin{equation}
	\sum_{i=1}^{N} \frac{\sum_{j={\rm after}}^{M_i} 
	\frac{R_{i,j}-\mu_i-\hat{s_0}}{\sigma_{i,j}^2}}{ \frac{1}{M_i} 
	\left[\sum_{j=1}^{\rm before } \left(\frac{R_{i,j}-\mu_{i} }{\sigma_{i,j} } 
	\right)^2
	+\sum_{j={\rm after}}^{M_i}  \left(\frac{R_{i,j}-\mu_{i} -\hat{s_0}}{ 
	\sigma_{i,j} } \right)^2\right]}=0\,.\label{eq:estis0}
\end{equation}
We can use a two-step iterative method to solve the MLE for $s_0$ from 
Equation~\ref{eq:estis0}, as follows: 
{\bf step i.} start with the initial value of 
$s_0=0$ and compute the re-scaled errorbar using \begin{equation}
	\sigma_{i,j}'^2=\sigma_{i,j}^2\frac{1}{M_i} \left[\sum_{j=1}^{\rm before } 
	\left(\frac{R_{i,j}-\mu_{i} }{\sigma_{i,j} } \right)^2
	+\sum_{j={\rm after}}^{M_i}  \left(\frac{R_{i,j}-\mu_{i} -\hat{s_0}}{ 
	\sigma_{i,j} } \right)^2\right]\,,
	\label{eq:resigma}
\end{equation} {\bf step ii.}  update value of $s_0$ using \begin{equation}
	\hat{s_0}'= \frac{\sum_{i=1}^{N} \sum_{j={\rm after}}^{M_i} 
	\left({R_{i,j}-\mu_i}\right) \sigma_{i,j}'^{-2}}{ \sum_{i=1}^{N} \sum_{j={\rm 
	after}}^{M_i} \sigma_{i,j}'^{-2}}\,.
	\label{eq:updat2}
\end{equation}
Here, Equation~\ref{eq:resigma} acts to estimate the errorbars of each pulsar 
using the total standard deviation after removing the jump.  
Equation~\ref{eq:updat2} uses the re-scaled errorbar to compute the 
error-weighted average of the jump.
The experiment shows that 3 to 5 iterations are usually enough to get the 
$\sigma_{i,j}'$ stable to within 15 digits with the our updating scheme.
The above results (Equation~\ref{eq:estimu} and \ref{eq:estis0}) agree with the 
ad-hoc $\chi^2=1$ approach, which is a rather common method to deal with the 
systematics in errorbar \cite{NR}.  

We now proceed to compute the error of the MLEs. From Equation~\ref{eq:updat2}, 
one can show that the variance of the estimator $\hat{s_0}$ has two parts
\begin{equation}
	{\rm Var}[s_0]\equiv \langle \hat{s_0}^2 \rangle-\langle \hat{s_0} \rangle^2 
	\simeq A+B\,,
	\label{eq:stds0}
\end{equation}
where the part $A$ and $B$ are
\begin{eqnarray}
	A&=&\frac{\sum_{i=1}^{N}
	\left(\sum_{j={\rm after}}^{M_{i}} \sigma_{i,j}^{-2}\right)^2 /\left( 
	\sum_{j=1}^{\rm before} \sigma_{i,j}^{-2}\right)}{\left( \sum_{i=1}^{N} 
	\sum_{j={\rm after}}^{M_i} \sigma_{i,j}^{-2}\right)^2}\,,\\
	B&=&\frac{1}{\sum_{i=1}^{N} \sum_{j={\rm after}}^{M_i} \sigma_{i,j}^{-2}}\,.
\end{eqnarray}
Part $A$ comes from the error in estimating the local $\mu_i$ using data 
before the jump, and part $B$ originates from the error of the data after the jump.  
The error of $\hat{s_0}$ is $\delta \hat{s_0}=\sqrt{  {\rm Var}[s_0]}$.  From 
Equation~\ref{eq:stds0}, one can see that the final error in the estimator agrees 
with the optimal weighting averaging. It is clear that we have used all the 
available information in the data when estimating the clock jitter amplitude. 

The above MLE, $\hat{s_0}$ and $\hat{\mu_i}$, are the estimators for the clock 
jump amplitude and residual leveling \emph{given a pre-determined clock jump 
epoch}. In practical situations, we will not know the clock jump epoch 
beforehand, and
we need to search for it. As shown in the next section, we can simply calculate 
$\hat{s_0}$ on a dense grid of clock-jump epoch trials. The most likely jump is 
the largest among the trials.

While in this discussion we focus on the clock jump signal,
in practice one also needs to 
account the pulsar timing-parameters fitting. One can implement the timing-model 
fitting by replacing $\mu_i$ in Equation~\ref{eq:likred2} with the 
linearised timing waveform. In this paper, we take a simpler iterative approach, 
where we repeat the step of fitting for the clock jump and for timing 
parameters several times, until the result converges. Note that in each iteration, 
we subtract the fitted jump from the site arrival times (TOAs at the observatory) 
when fitting the timing parameters.
Such an iterative method also helps to derive the clock jump referring 
to the telescope site.

\section{Example using pulsar timing data from the Yunnan observatory}\label{sec:3}

\begin{figure}[H]
	\includegraphics[width=3.7in]{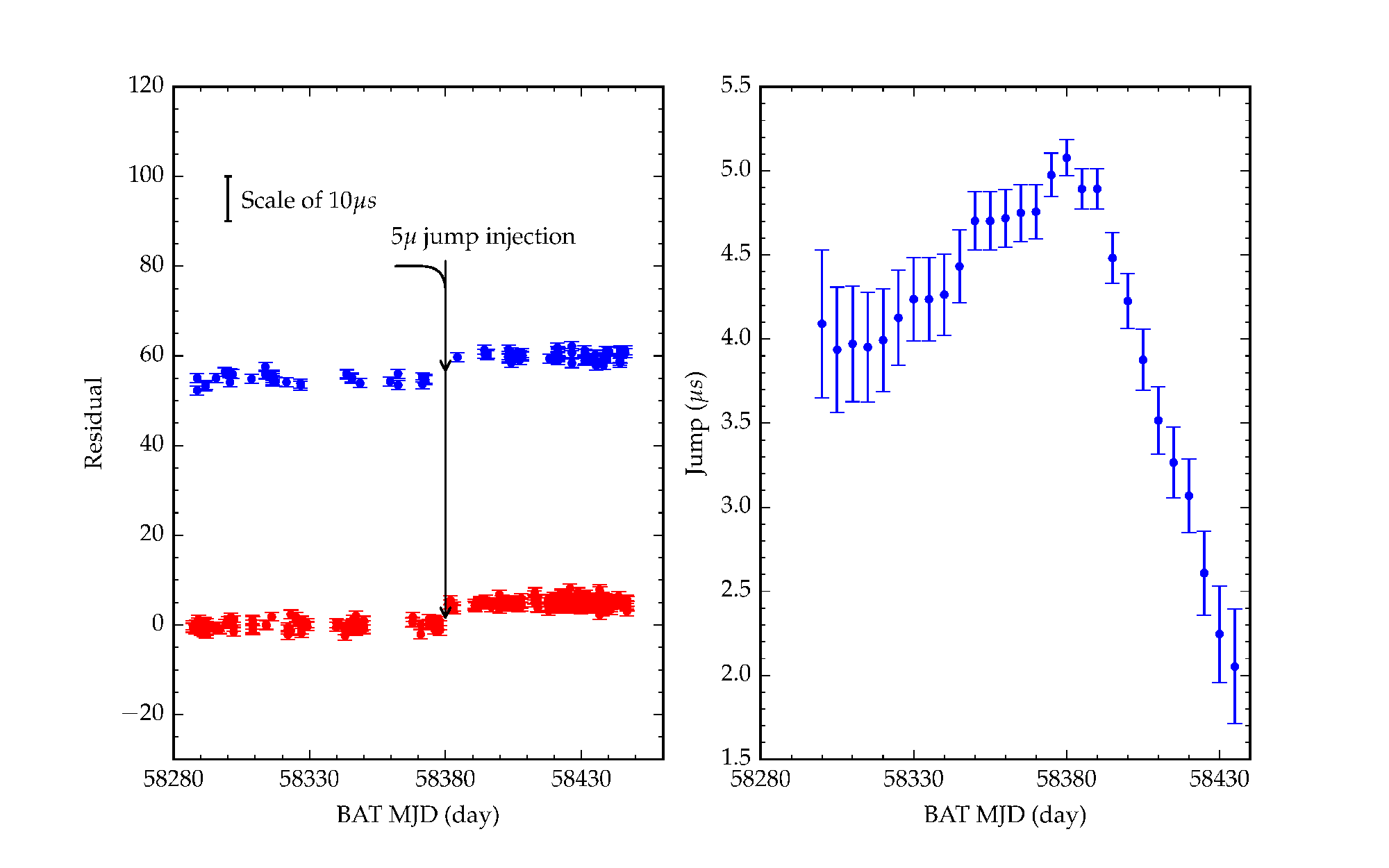}
	\caption{Left panel: The simulated pulsar (relative)timing residual for two pulsars. The 
	x-axis is the barycentric TOA in Modified Julian Day (MJD) and the y-axis is 
	units of time.  The scale is drawn and labeled on the top-left corner of the 
	figure.   We injected a 5-$\mu$s clock jump at MJD 58380.  Right panel: The 
	recovered clock jump (y-axis) as a function of jump epoch (x-axis).  One can 
	see we recover the clock jump successfully, where the
	measured jump is $5.08\pm 0.1$ $\mu$s.  \label{fig:sim}}
\end{figure}

We first validate our algorithm using simulated data. We simulate 
timing data of two pulsars (PSRs J1713+0747 and J0437-4715, which are the 
most accurately timed pulsars in our campaign) with similar length and cadence as the real data, which 
we are going to discuss right after, in this section. The data is simulated with precision 
of 5 $\mu$s. We inject a clock jump at MJD 58380 with an amplitude of $5$ $\mu$s.  
The data and recovered clock jump value are shown in Figure~\ref{fig:sim}. As 
one can see, the current algorithm recovers the clock jump amplitude well. 

We now turn to the real data. The \emph{Yunnan Astronomical Observatory} (YNAO) 
of the Chinese Academy of Science operates the Kunming 40-meter radio telescope 
(KM40m). The telescope was built in 2006
for the Chinese lunar-probe mission. It is located in the south-west
of China (${\rm N} 25^\circ01'38''$, ${\rm E} 102^\circ47'45''$),
approximately 15 kilometers away from the city of Kunming. The
total collecting area of KM40m is 1250 m$^2$. Our timing data were
collected using a room-temperature receiver (70 K system temperature), originally designed
for communication purposes. The center frequency is 2.5 GHz and 
the bandwidth is 300 MHz.

We have been regularly timing five millisecond pulsars since 2017. The observing 
schedule is much more intense compared to a typical pulsar timing program (e.g.  
see \cite{DCL16}), as we observe the same set of pulsars almost daily.  Part 
of our timing efforts aim at helping to identify the potential issues in the 
frequency-clock signal chain of telescope systems. Following the standard timing 
pipeline, we measured the TOAs using the software \textsc{PSRCHIVE} 
\cite{HVM04}, and computed the timing residual using \textsc{TEMPO2} 
\cite{EHM06,HEM06b}.

\begin{figure}[H]
	\includegraphics[width=3.6in]{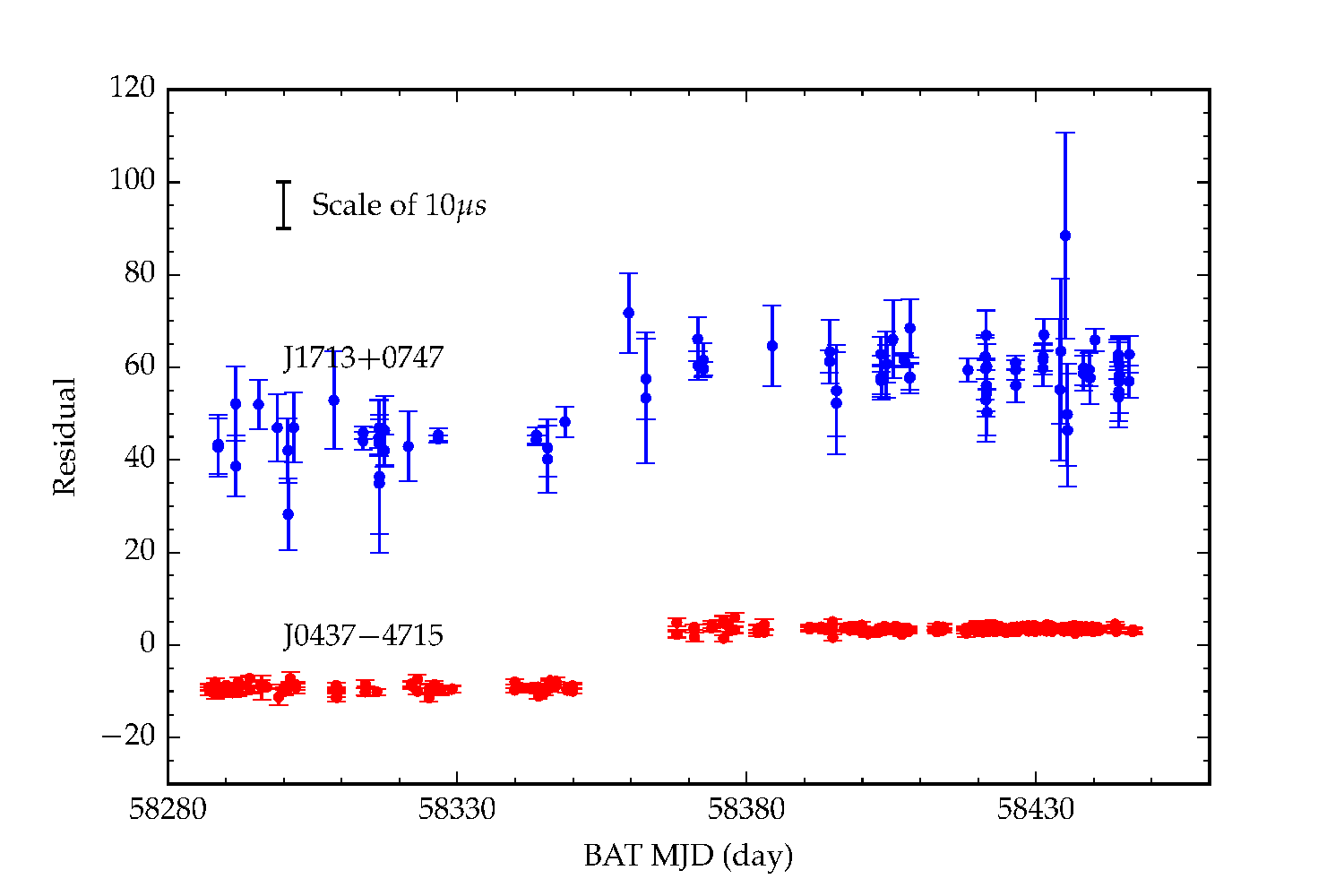}
	\caption{The timing residuals of the two most accurate millisecond pulsars in 
	our timing array pool, i.e. PSR J1713$+$0747 and PSR J0437$-$4715.  
	The x-axis is the barycentric TOA in Modified Julian Day (MJD) and the y-axis 
	is the time. The scale is drawn and labeled on the top-left corner of the 
	figure. Since the timing dataset have arbitrary offsets, we can simply compare the 
	two pulsars in the same panel. There is a clock jitter around MJD 58355, i.e.  
	25/08/2018, because of the signal path switching.  \label{fig:dat}}
\end{figure}

On the 25th of August 2018, the clock signal distributor was broken during an
intense lightning storm. Switching to the back-up signal path led to a jump 
in the pulsar timing signal, as shown in Figure~\ref{fig:dat}. 

\begin{figure}[H]
	\includegraphics[width=3.2in]{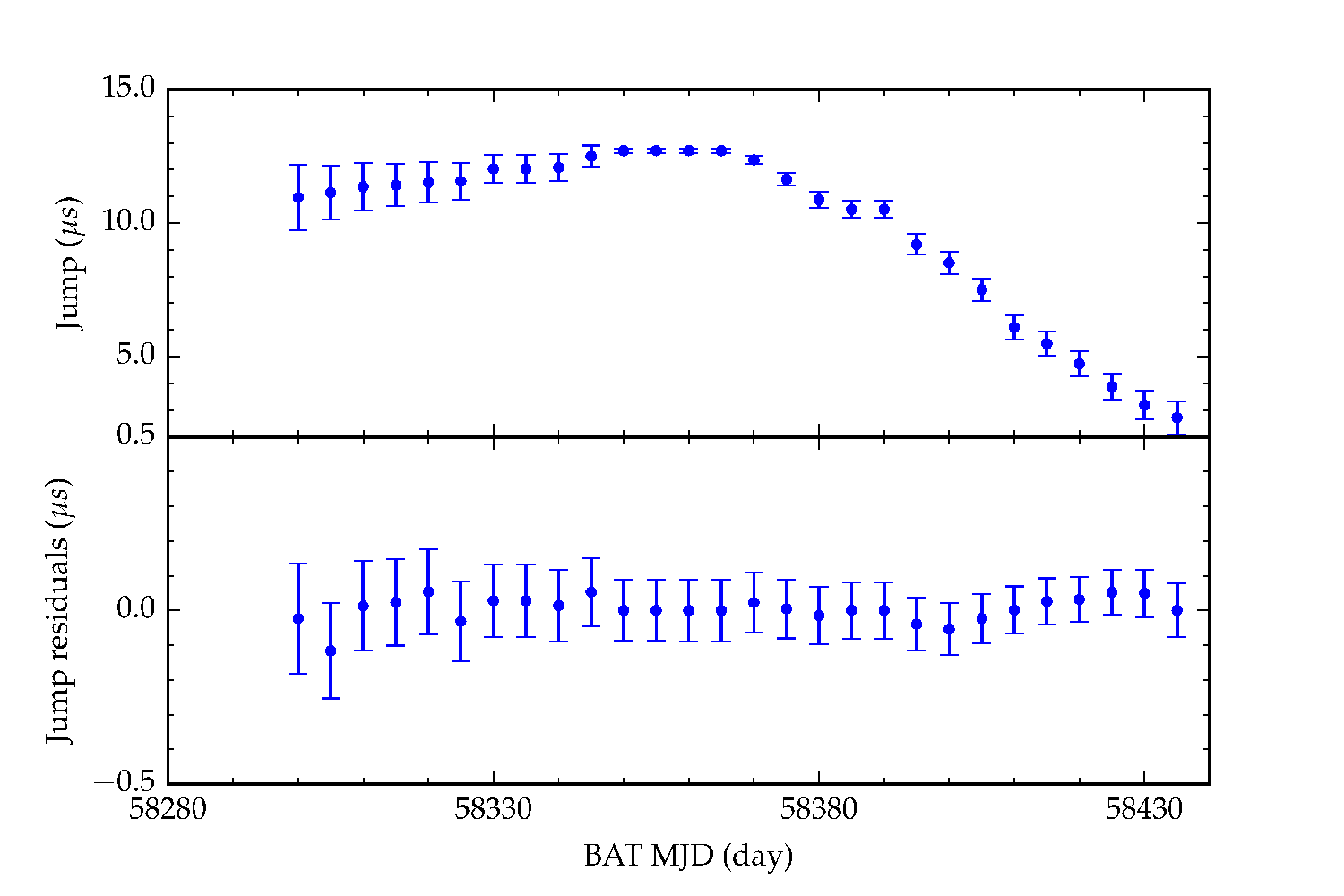}
	\caption{Upper panel: The inferred jumps as the function of jump time. The 
	x-axis is the time in MJD, the y-axis is the amplitude of the clock jump. The 
	inferred clock jump peaks between MJD 58350 and 58365, the amplitude is 
	$12.71\pm0.08$ $\mu$s. The error-bar is the standard deviation of the 
	estimator. Note that there is a plateau from MJD 58350 to 58365. This is 
	due to the data gap, in which time interval we can only deduce the time window of the clock jump, 
	rather than pin down the exact epoch of clock jump. Lower panel: The inferred 
	jumps after removing the 12.71 $\mu$s jump at MJD 58350.  No jumps can be 
	further detected with more than 1-$\sigma$ confidence.
	\label{fig:jmp}}
\end{figure}

The search of the clock jump is shown in Figure~\ref{fig:jmp}, where we compute 
the clock-jump value on a grid of jump epoch.  The maximum clock jump value 
appears in the span between MJD 58350 and MJD 58365. After correcting the jump 
at MJD 58350, no significant clock jump can  be further detected. The measured 
jump is $12.71\pm0.08$ $\mu$s. We have determined the clock jump with a
precision of 80 ns, which is 8 times lower than the original pulsar timing 
precision (600 ns for PSR 0437-4715). The 80-ns precision mainly comes from the 
statistical algorithm presented in this paper.

\section{Conclusion and Discussion}
\label{sec:discon}

In this work, we have investigated a computationally efficient method to measure the observatory's local 
clock jumps using pulsar timing data as the first step to build the autonomous 
pulsar time-scale. In order to minimize errors in determining the clock jump, we 
used Bayesian statistics, with which the unknown systematics can be 
marginalized with an analytic calculation. The estimator we adopted is the MLE, 
which is thought to be asymptotically optimal. We found that the Gaussian 
likelihood after marginalization over the noise scales leads to a distribution 
similar to the Student's $t$-distribution. Using the method, we measured the 
clock jump at Yunnan observatory using the pulsar timing data, where 80-ns 
precision could be achieved using the observed data. The current precision is 
worse than that of widely used GPS time delivery service. Using a better receiver 
(e.g. 20 K noise temperature) with wider observation bandwidth (e.g. 900MHz) 
will bring the precision down to an 8-ns level, which will be better than that of the 
current GPS time delivery service. In addition, the precision can also be improved 
if some larger facilities can be used, for example, the Five-hundred-meter
Aperture Spherical radio Telescope (FAST).
FAST has just complete its initial commissioning \cite{JP19} and first studies on pulsars have also been made by using it \cite{LJG19a, Yu19, LJG19b, WHF19, QL19}.
Moreover, in general, the clock precision scales as 
$\sim N_{\rm psr}^{1/2}$, with $N_{\rm psr}$ the number of pulsar observed.  
If more telescope time is available, we can time more pulsars, which would allow 
substantial improvement on the precision of pulsar time-scales. 
 In this way, 
pulsar timing may be used as an independent method to check the local clock 
stability, if the GPS signal is not available or if the autonomy is the 
priority.

Due to the intrinsic phase confusion of the pulsar timing technique, the clock-jump 
value ($s$) seen in the timing residuals is the remainder of the division of the real 
clock jump ($s_{\rm real}$) by pulsar period $p$, i.e. $s_{\rm real}= s\, ({\rm 
mod\,} p)$. In this way, one will not recover a clock jump that is larger than the 
rotational period of the pulsar in the array with the longest one. 
This difficultly can be resolved by observing a long period 
pulsar in the array. The long-period pulsar can help to derive an initial 
guess of the jump value. For our analysis, we used PSR B0329+54, which has
a rotational period of 0.7\,s. The MSP timing data can be preconditioned to the 
coherent state, and the final clock jump without phase confusion can be measured 
with the MSP ensemble.

The measured clock jump in the timing residuals refers to the
barycentric dynamical time. However,
in practice we need the clock jump with respect
to the \emph{local} terrestrial time (most of the times, Coordinated
Universal Time). In this paper, we perform the reference frame correction
using an iterative approach, where we can measure the clock jump
in the barycentric frame, correct the local TOA with the measured
jump, reform the timing residual (with timing-parameter fitting), and 
repeat the iteration. Once
the process converges, i.e. when no further clock correction is needed,
one determines the clock jump in the terrestrial frame. The iterative method 
thus also enables us to fit for the pulsar timing parameters.

In the current paper, we focus on the clock jump, so the timing-parameter 
modeling is not explicitly shown. One can include the timing parameter modeling 
in the Equation~\ref{eq:likred2} by replacing the central value $\mu_i$ with the 
linearised timing model \cite{HEM06b}. Here, the iterative procedure 
automatically takes care of the timing parameter fitting.

Pulsar timing residuals often contain red noise, i.e. time-correlated noise. In the 
current paper, we consider white noise only, which reduces the computational 
cost significantly. There are two major reasons that legitimize our decision to take such a 
simplified approach. Firstly, the clock jump is highly localized in time, 
therefore we do not need an extended dataset to measure its value. 
The contribution of red noise in such a short time scale is small and usually the white noise dominates. 
Secondly, we have marginalized the `Efac' parameters, which take care of the noise systematics.

 In order to get a robust error estimation, we include the `EFac' in our 
model. We neglected the jitter noise modeling because the radiometer noise 
(error bars) is much larger than the jitter level. For example, the 1-hr 
weighted precision of J0437$-$4715 is 600 ns at YN40m, which is 6-20 times  
larger than the jitter noise measured \cite{LKL12}. Similarly, the current 
accuracy for the clock jump is 80 ns based on 150-day observation.  The jitter 
noise contribution is further averaged out by another order of magnitude with 
the current data processing algorithm.

{\it This work was supported by XDB23010200, 11690024, NSFC U15311243, 
2015CB857101 and funding from TianShanChuangXinTuanDui and the Max-Planck 
Partner Group. }


\begin{thebibliography}{99}
\bibitem{HL11}
J.~G. {Hartnett}, and A.~N. {Luiten}, \href{https://doi.org/10.1103/RevModPhys.83.1}\revmp ~83, 1 (2011).

\bibitem{MT97}
D.~N. {Matsakis}, J.~H. {Taylor}, and T.~M. {Eubanks},  \href{http://adsabs.harvard.edu/abs/1997A\%26A...326..924M}\ana ~326, 924 (1997).

\bibitem{HCM12}
G. {Hobbs}, W. {Coles}, R.~N. {Manchester}, M.~J. {Keith},
  R.~M. {Shannon}, D. {Chen}, M. {Bailes}, N.~D.~R.{Bhat}, S. {Burke-Spolaor}, D. {Champion},
 A. {Chaudhary}, A. {Hotan}, J. {Khoo}, J. {Kocz}, Y. {Levin}, S. {Oslowski}, B. {Preisig}, V. {Ravi},
  J.~E. {Reynolds}, J. {Sarkissian}, W. {van Straten}, J.~P.~W. {Verbiest}, D. {Yardley}, and X.~P. {You} 
\href{https://doi.org/10.1111/j.1365-2966.2012.21946.x}\mnras ~427, 2780 (2012).

\bibitem{RF18}
A.~E. {Rodin}, and V.~A. {Fedorova}, 
\href{https://doi.org/10.1134/S1063772918060057}\astrp ~62, 378 (2018).

\bibitem{GP11}
B. {Guinot}, and G. {Petit}, 
\href{http://adsabs.harvard.edu/abs/1991A\%26A...248..292G}\ana ~248, 292 (1991).

\bibitem{R08}
A.~E. {Rodin}, \href{https://doi.org/10.1111/j.1365-2966.2008.13270.x}\mnras ~387, 1583 (2008).

\bibitem{CLL16} R.~N. Caballero, K.~J. Lee, L. Lentati, G. Desvignes,
  D.~J. Champion, J.~P.~W. Verbiest, G.~H. Janssen, B.~W. Stappers, M. Kramer, P. Lazarus,
  A. Possenti, C. Tiburzi, D. Perrodin, S. Os{\l}owski, S. Babak, C.~G. Bassa, P. Brem,
  M. Burgay, I. Cognard, J.~R. Gair, E. Graikou, L. Guillemot, J.~W.~T. Hessels,
  R. Karuppusamy, A. Lassus, K. Liu, J. McKee, C.~M.~F. Mingarelli, A. Petiteau, M.~B. Purver,
  P.~A. Rosado, S. Sanidas, A. Sesana, G. Shaifullah, R. Smits, S.~R. Taylor, G. Theureau,
  R. van Haasteren, and A. Vecchio, \href{https://doi.org/10.1093/mnras/stw179}\mnras ~457, 4421 (2016).

\bibitem{kaplan05} E. Kaplan, and C. Hegarty, {\it Understanding GPS: principles and applications} (Artech house, Boston, USA, 2005).

\bibitem{mic2015}
Microsemi, {\it Datasheet of SyncSystem 4380A Master Timing Reference}
  (Microsemi Corporate Headquarters, One Enterprise, Aliso Viejo, CA 92656 USA, 2015).

\bibitem{LKG16} P. Lazarus, R. Karuppusamy, E. Graikou, 
R.~N. Caballero, D.~J. Champion, K.~J. Lee, J.~P.~W. Verbiest, 
and M. Kramer, \href{https://doi.org/10.1093/mnras/stw189}\mnras ~458, 868 (2016).

\bibitem{FB90} R.~S. {Foster}, and D.~C. {Backer}, \href{https://doi.org/10.1086/169195}\apj ~361, 300 (1990).

\bibitem{LK12}
D.~R. {Lorimer} and M. {Kramer}, {\it Handbook of Pulsar Astronomy}
  (Cambridge University Press, Cambridge, UK, 2012).

\bibitem{wilks1938large}
S.~S. Wilks, \annms ~9, 60 (1938).

\bibitem{HEM06b}
G.~B. {Hobbs}, R.~T. {Edwards}, and R.~N. {Manchester}, 
\href{https://doi.org/10.1111/j.1365-2966.2006.10302.x}\mnras ~369, 655 (2006).

\bibitem{gregory2005bayesian} P. Gregory, {\it Bayesian Logical Data Analysis for the Physical Sciences: A
  Comparative Approach with Mathematica {\textregistered} Support} (Cambridge University Press, Cambridge, UK, 2005).
  
\bibitem{gr14}
I.~S. Gradshteyn, and I.~M. Ryzhik, \textit{Table of integrals, series, and products} (Academic press, London, UK, 2014).  

\bibitem{NR} W.~H. Press, S.~A. Teukolsky, W.~T. Vetterling, and B.~P. Flannery,
 {\it Numerical recipes in C}, Vol.~2 (Cambridge university press, Cambridge, UK, 1982).

\bibitem{DCL16} G. Desvignes, R.~N. Caballero, L. Lentati,
  J.~P.~W. Verbiest, D.~J. Champion, B.~W. Stappers, G.~H. Janssen, P. Lazarus, S. Os{\l}owski,
  S. Babak, C.~G. Bassa, P. Brem, M. Burgay, I. Cognard, J.~R. Gair, E. Graikou, L. Guillemot, J.~W.~T. Hessels, A. Jessner, C. Jordan, R. Karuppusamy, M. Kramer,
  A. Lassus, K. Lazaridis, K.~J. Lee, K. Liu, A.~G. Lyne, J. McKee, C.~M.~F. Mingarelli,
  D. Perrodin, A. Petiteau, A. Possenti, M.~B. Purver, P.~A. Rosado, S. Sanidas, A. Sesana,
  G. Shaifullah, R. Smits, S.~R. Taylor, G. Theureau, {Tiburzi}, R. van Haasteren, and A. Vecchio, 
  \href{https://doi.org/10.1093/mnras/stw483}\mnras ~ 458, 3341 (2016).
  
\bibitem{HVM04}
A.~W. {Hotan}, W. {van Straten}, and R.~N. {Manchester}, \href{https://doi.org/10.1071/AS04022}\pasa ~21, 302 (2004).

\bibitem{EHM06} R.~T. {Edwards}, G.~B. {Hobbs}, and  R.~N. {Manchester}, \href{https://doi.org/10.1111/j.1365-2966.2006.10870.x}\mnras ~372, 1549 (2006).

\bibitem{JP19}  P. Jiang, Y.~L. Yue, H.~Q. Gan, R. Yao, H. Li,
G.~F. Pan, J.~H. Sun, D.~J. Yu, H>~F. Liu, N.~Y. Tang, 
L. Qian, J.~G. Lu, J. Yan, B. Peng, S.~X. Zhang, Q.~M. Wang,
Q. Li, D. Li, and FAST Collaboration,
  \href{https://doi.org/10.1007/s11433-018-9376-1}\scich ~62, 959502 (2019).

\bibitem{LJG19a} J.~G Lu, B. Peng, K. Liu, P. Jiang, Y.~L. Yue,
M. Yu, Y.~Z. Yu, F.~F. Kou, L. Wang, and FAST Collaboration \href{https://doi.org/10.1007/s11433-018-9372-7}\scich
~62, 959503 (2019).

\bibitem{Yu19} Y.~Z. Yu, B. Peng, K. Liu, C.~M. Zhang, L. Wang,
F.~F. Kou, J.~G. Lu, M. Yu, and FAST Collaboration, 
\href{https://doi.org/10.1007/s11433-018-9358-8}\scich
~62, 959504 (2019).

\bibitem{LJG19b} J.~G Lu, B. Peng, R.~X. Xu, M. Yu, S. Dai, W.~W. Zhu, 
Y.~Z. Yu, P. Jiang, Y.~L. Yue, L. Wang, and FAST Collaboration
 \href{https://doi.org/10.1007/s11433-019-9394-x}\scich
~62, 959505 (2019).

\bibitem{WHF19} H.~F. Wang, W.~W. Zhu, P. Guo, D. Li, S.~B. Feng,
Q. Yin, C.~C. Miao, Z.~Z. Tao, Z.~C. Pan, P. Wang, X. Zheng,
X.~D. Deng, Z.~J. Liu, X.~Y. Xie, X.~H. Yu, S.~P. You,
H. Zhang, and FAST Collaboration, 
\href{https://doi.org/10.1007/s11433-018-9388-3}\scich
~62, 959507 (2019).

\bibitem{QL19} L. Qian, Z.~C. Pan, D. Li, G. Hobbs, W.~W. Zhu, 
P. Wang, Z.~J. Liu, Y.~L. Yue, Y. Zhu, H.~F. Liu, D.~J. Yu,
J.~H. Sun, P. Jiang, G.~F. Pan, H. Li, H.~Q. Gan, R. Yao,
X.~Y. Xie, F. Camilo, A. Cameron, L. Zhang, S. Wang, and FAST Collaboration, \href{https://doi.org/10.1007/s11433-018-9354-y}\scich
~62, 959508 (2019).

\bibitem{LKL12} K. {Liu}, E.F. Keane, K.J. {Lee}, M. Kramer, J.M. Cordes, and 
	M.B.  {Purver}, \href{https://doi.org/10.1111/j.1365-2966.2011.20041.x}\mnras ~420, 361 (2012).  
	
\end{thebibliography}

\begin{appendix}




%
%
%
%
%
\end{appendix}

\end{multicols}
\end{document}